# Wavelength-Tunable Infrared Metamaterial by Tailoring Magnetic Resonance Condition with VO$_2$ Phase Transition

Hao Wang, Yue Yang, and Liping Wang[*]

*School for Engineering of Matter, Transport & Energy, Arizona State University, Tempe, Arizona, 85287 USA*

**\*** *Corresponding author: liping.wang@asu.edu*

In this work, we report the design of a wavelength-tunable infrared metamaterial by exciting magnetic resonance with phase transition of vanadium dioxide (VO$_2$). Numerical simulation shows a broad absorption peak at the wavelength of 10.9 μm when VO$_2$ is a metal, but it shifts to 15.1 μm when VO$_2$ changes to dielectric phase below its phase transition temperature of 68°C. The large tunability of 38.5% in the resonance wavelength stems from the different excitation conditions of magnetic resonance assisted by plasmon in metallic VO$_2$ but optical phonons in dielectric VO$_2$. The physical mechanism is elucidated with the aid of electromagnetic field distribution at the resonance wavelengths. A hybrid magnetic resonance mode due to plasmon-phonon coupling is also discussed. The results here would be beneficial for active control in novel electronic, optical and thermal devices.

The unique phase transition behavior of vanadium dioxide (VO$_2$) [1-3] has drawn lots of attentions recently and many applications have been found. Optical properties of VO$_2$ changes dramatically when phase transition from dielectric to metal occurs at 68°C, which can be thermally induced by temperature control. Applications of VO$_2$ have been demonstrated in optical information storage [4], strain sensing [5], and lithium-ion batteries [6]. Moreover, dielectric VO$_2$ possesses several optical phonon modes in the infrared, which has been employed to modulate radiative heat transfer [7] in designing novel thermal devices such as thermal diodes/rectifiers [8,9] and thermal transistors [10].

Efforts have been made recently in tunable metamaterials made of phase transition VO$_2$. Dicken et al. [11] demonstrated a resonance-frequency-tunable metamaterial by depositing split-ring resonators on a VO$_2$ film. Kats et al. [12] showed ~10% resonance wavelength tunability with a plasmonic antenna array on a VO$_2$ film in the mid-infrared. They also demonstrated an ultra-thin tunable perfect absorber based on the VO$_2$ phase transition [13]. Previous designs were realized by modulating the excitation condition of surface plasmon polariton (SPP) at the interface of subwavelength plasmonic nanostructures and the supporting VO$_2$ film upon phase transition. However, the tunable frequency range was quite limited from the reported work.

Magnetic resonance has been studied intensively for designing selective thermal emitters [14,15] and perfect metamaterial absorbers [16]. Magnetic resonance occurs when external electromagnetic wave couples with magnetic resonance excited inside the metamaterial structures typically in a metal-insulator-metal configuration, resulting in strong absorption or emission at the resonance frequency. Note that, magnetic resonance can be also excited in the mid-infrared regime with polar materials, mediated by optical phonons, rather than plasmon in metals. An infrared selective emitter made of SiC has been demonstrated by exciting magnetic resonance within its phonon absorption band [17]. The induced electrical current by the resonant magnetic field is realized by the high-frequency vibration of ions in SiC, rather than free charges in metals.

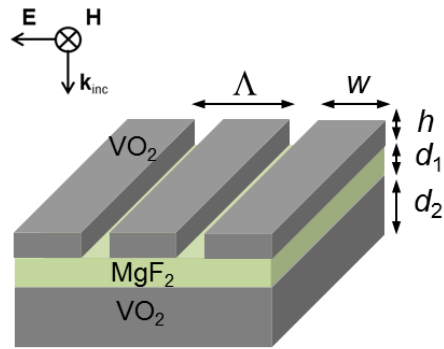

Fig. 1. Proposed 1D tunable metamaterial structure with period $\Lambda$ = 1.5 μm, strip width $w$ = 1.25 μm, layer thicknesses $h$ = 0.5 μm, $d_1$ = 0.3 μm, and $d_2$ = 1 μm. The phase transition of VO$_2$ can be controlled by modulating the temperature.

In this study, we present an infrared metamaterial by exciting magnetic resonance at different conditions with either metallic or dielectric VO$_2$, leading to highly tunable resonant wavelength upon the phase transition of VO$_2$. Fig. 1 depicts the proposed tunable metamaterial structure, which is made of a one-dimensional VO$_2$ periodic grating structure (period $\Lambda$ = 1.5 μm and strip width $w$ = 1.25 μm) on stacked MgF$_2$ and VO$_2$ film. The thicknesses of the VO$_2$ grating and thin films are $h$ = 0.5 μm, $d_1$ = 0.3 μm, and $d_2$ = 1 μm. The temperature of the structure can be modulated to thermally control VO$_2$ phase transition.

When the temperature is above 68°C, VO$_2$ is an isotropic metal, whose electrical permittivity $\varepsilon_m$ can be described by a Drude model as [1]

$$\varepsilon_m = -\varepsilon_\infty \frac{\omega_p^2}{\omega^2 + i\omega\Gamma} \quad (1)$$

where $\omega$ is angular frequency, $\varepsilon_\infty$ = 9 is the high-frequency constant, $\omega_p$ = 8000 cm$^{-1}$ is the plasma frequency, and $\Gamma$ = 10000 cm$^{-1}$ is the collision frequency. When the temperature is below 68°C, VO$_2$ becomes dielectric but with uniaxial anisotropy. Considering (200)-oriented crystal VO$_2$ with optical axis normal to the surface [1], it exhibits ordinary dielectric response denoted as $\varepsilon_O$ when incident electric field is perpendicular to optical axis, and extraordinary response $\varepsilon_E$ when electric field is parallel to optical axis. Both components can be described by a classical oscillator model as

$$\varepsilon(\omega) = \varepsilon_\infty + \sum_{j=1}^{N} \frac{S_j \omega_j^2}{\omega_j^2 - i\gamma_j\omega - \omega^2} \quad (2)$$

where $\omega_j$ is the phonon vibration frequency, $\gamma_j$ is the scattering rate, $S_j$ represents the oscillation strength, and $j$ is the phonon mode index. The values for each parameter can be found from [1] for both ordinary ($\varepsilon_O$) and extraordinary ($\varepsilon_E$) components with a total of eight phonon modes for $\varepsilon_O$ and nine modes for $\varepsilon_E$. In the simulation, the permittivity tensor was employed to consider the uniaxial anisotropy of dielectric VO$_2$:

$$\bar{\bar{\varepsilon}}_{\text{dielectric}} = \begin{pmatrix} \varepsilon_O & 0 & 0 \\ 0 & \varepsilon_O & 0 \\ 0 & 0 & \varepsilon_E \end{pmatrix} \quad (3)$$

Fig. 2(a) plots the real parts for the permittivity of metallic and dielectric VO$_2$ in the mid-infrared regime from 5 μm to 20 μm in wavelength. The metallic phase exhibits a negative real part of permittivity, which is crucial to excite plasmonic resonances as found in most noble metals [18]. On the other hand, there exist several phonon modes in both ordinary and extraordinary components of the dielectric phase. As a result, negative real part of permittivity exists in the wavelengths from 12.4 μm to 16.7 μm for $\varepsilon_O$ and from 17.5 μm to 18.8 μm for $\varepsilon_E$, respectively. As pointed out by Ref. [17], negative permittivity is required to excite phonon-mediated magnetic resonance in polar materials. The unique material properties of metallic and dielectric VO$_2$ suggest the potential in exciting magnetic resonance in both phases.

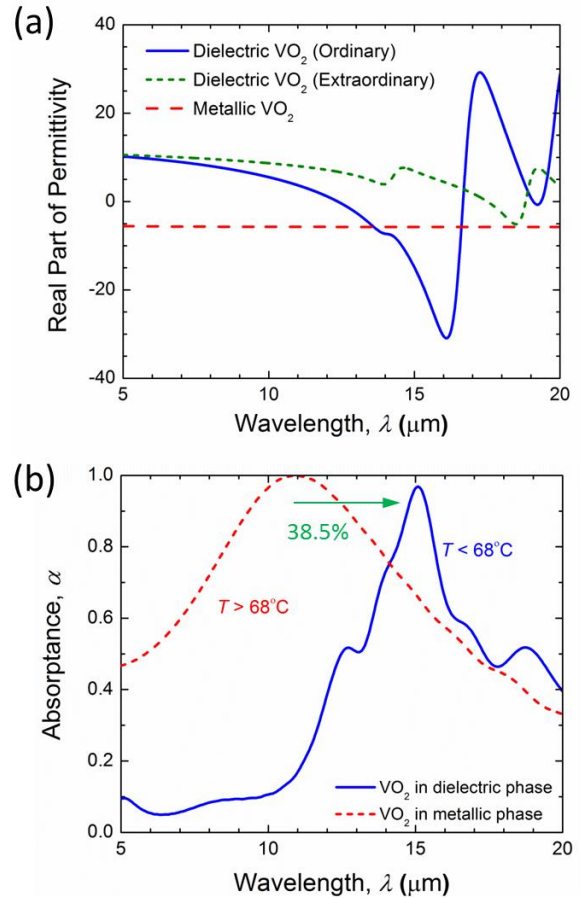

Fig. 2. (a) Real parts of permittivity of VO$_2$ at different phases. (b) Simulated normal absorptance of proposed tunable metamaterial in the mid-infrared upon VO$_2$ phase transition, showing a relative 38.5% shift of resonant absorption peak wavelength.

The finite-difference time-domain method (Lumerical Solutions, Inc.) was used to numerically calculate the spectral reflectance $R$ and transmittance $T$ of the proposed tunable metamaterial above and below the VO$_2$ phase transition temperature of 68°C. The optical constants of MgF$_2$ were obtained from Palik's data [19]. A linearly polarized plane wave was incident normally onto the metamaterial structure with transverse-magnetic (TM) waves, in which magnetic field is along the grating groove direction. Note that, magnetic resonance can be excited only at TM polarization in 1D grating based metamaterials [14,15]. A numerical error less than 2% was verified with sufficiently fine mesh sizes.

As shown in Fig. 2(b), the normal absorptance in the infrared region from 5 μm to 20 μm was thus obtained by $\alpha = 1-R-T$ based on energy balance. When the temperature is above 68°C, the VO$_2$ is at metallic phase and the metamaterial exhibits a broad absorption band peaked at the wavelength of 10.9 μm with almost 100% absorption. However, when VO$_2$ becomes dielectric at temperatures less than 68°C, the absorption band is narrower and shifts to the peak wavelength of 15.1 μm with maximum absorptance of 0.97, resulting in a relative 38.5% peak wavelength shift upon the $VO_2$ phase transition from metal to dielectric induced thermally. Note that there exist two bumps on the shoulder of absorption peak around 16.5 μm and 19 μm, which are caused by the abrupt changes of optical properties of dielectric $VO_2$ due to strong phonon vibration.

In fact, both absorption peaks are caused by the excitations of magnetic resonance at both phases of $VO_2$. But the fundamental difference is that, one is assisted by free charges or plasmon in metallic phase, while the other is mediated by optical phonons in its dielectric phase. The different resonance conditions and thus the resulting large resonance wavelength shift are due to different optical behaviors of different energy carriers that excite the magnetic resonances.

To illustrate the underlying mechanism responsible for the large absorption peak shift, electromagnetic field distributions at the cross section of the metamaterial structure were plotted at the resonance wavelengths with metallic and dielectric phases of $VO_2$, as shown in Figs. 3(a) and 3(b), respectively. The arrows indicate the strength and direction of the electric field vectors, while the contour shows magnetic field strength normalized to the incidence as $|H/H_0|^2$ at different locations.

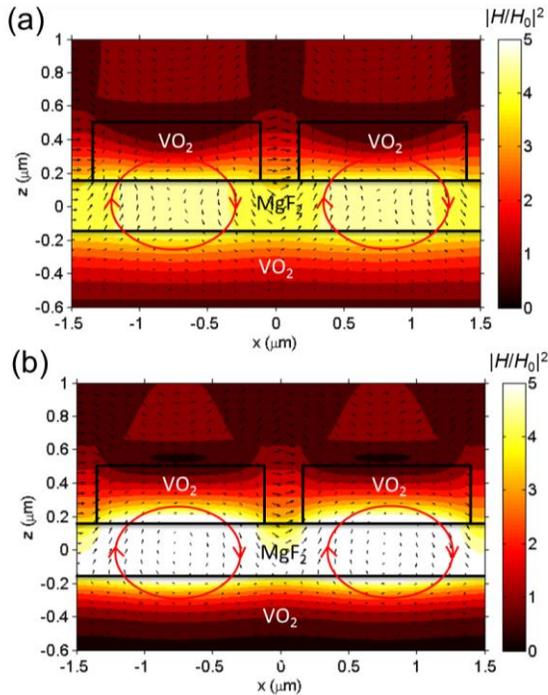

Fig. 3. Electromagnetic field distribution at resonance peak wavelengths when $VO_2$ is at (a) metallic or (b) dielectric phase. The field patterns show the exact behavior of magnetic resonance with both phases of $VO_2$, but assisted by plasmon in metallic $VO_2$ and mediated by optical phonons in dielectric $VO_2$, respectively.

When $VO_2$ is at metallic phase, the electric field vectors inside the $MgF_2$ layer underneath the $VO_2$ strips suggest an anti-parallel current loop, along with the strong localization of magnetic field, as shown in Fig. 3(a). The localized energy is more than five times higher than the incidence. This is the exact behavior of magnetic resonance that has been intensively studied in similar 1D grating based metamaterials [14,15]. Due to the oscillating movement of free charges in metallic $VO_2$, the sandwiched $MgF_2$ layer serves as a capacitor, while top metallic $VO_2$ strip and the bottom metallic $VO_2$ film function as inductors, forming a resonant alternating-current circuit. When the magnetic resonance occurs, the external electromagnetic energy at the resonant wavelength of 10.9 μm is coupled with the oscillating plasmon, resulting in almost 100% absorption inside the metamaterial structure.

When $VO_2$ becomes dielectric with the temperature below 68°C, the electromagnetic field shown in Fig. 3(b) presents a similar behavior of magnetic resonance with an induced anti-parallel electric current loop and confined magnetic field inside the $MgF_2$ layer but at a different resonance wavelength of 15.1 μm. The localized energy strength is about five times to the incidence. Note that, this resonant wavelength is within the phonon absorption band of the ordinary component of dielectric $VO_2$, in which negative permittivity exists. When optical phonons vibrate at high frequency, the fast movements of bound charges or ions form oscillating electric currents and an inductor-capacitor resonant circuit, resulting in the excitation of magnetic resonance. Since the energy carrier changes from free electrons to optical phonons upon the phase transition of $VO_2$ from metal to dielectric, a large shift in resonance wavelengths occurs. It should be noted that, similar to the surface phonon polariton with polar materials [20], which is a counterpart of SPP in the infrared regime, phonon-mediated magnetic resonance [17] is the counterpart of magnetic resonance in plasmonic metamaterials made of metallic nanostructures [14-16].

Finally, we would like to show that, a hybrid magnetic resonance mode could also occur by the phonon-plasmon coupling from a modified tunable metamaterial by replacing the bottom $VO_2$ layer with a gold film, as shown in the inset of Fig. 4. The period and strip width of the top $VO_2$ grating are kept unchanged, while the thicknesses of the grating and the $MgF_2$ spacer layer are $h = d_1 = 0.5$ μm.

The spectral normal absorptance of the hybrid structure is plotted in Fig. 4. The absorption peaks when $VO_2$ is in either metallic or dielectric phase remain almost at the same resonance wavelengths, suggesting that magnetic resonance can still be excited in both phases of $VO_2$. However, maximum absorptance drops slightly to 0.85 for the peak with metallic $VO_2$, while the absorption peak with dielectric $VO_2$ becomes narrower, after the bottom $VO_2$ film was replaced by a gold substrate. The successful excitation of magnetic resonance between metallic $VO_2$ strips and the bottom gold film is easy to understand, since there exist free charges in both metals. On the other hand, it would be expected that it would fail to excite phonon-mediated magnetic resonance due to the removal of the bottom $VO_2$ film. Surprisingly, the strong absorption with dielectric $VO_2$ could still occur. This can be understood by the excitation of a hybrid magnetic resonance mode due to the strong coupling between optical phonons in dielectric $VO_2$ and plasmon in the bottom gold substrate. The vibration of optical phonons at high frequency at the top interface of the $MgF_2$ spacer along with the movement of plasmon at the bottom interface could still form a close-loop inductor-capacitor circuit, which successfully excites magnetic resonance at the wavelength of 14.8 μm.

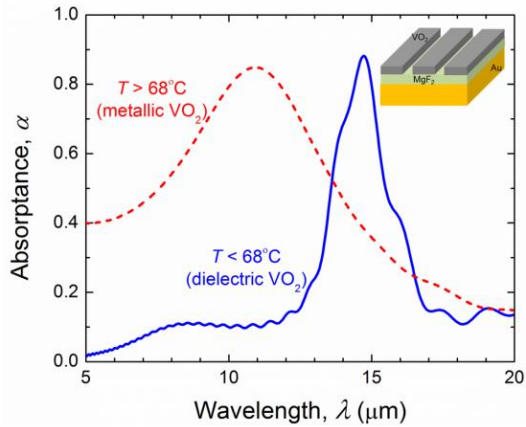

Fig. 4. Normal absorptance of the metamaterial with gold substrate instead of the bottom $VO_2$ layer. The absorption peaks still exists because of a hybrid magnetic resonance mode due to phonon-plasmon coupling between top dielectric $VO_2$ and bottom gold.

In conclusion, we have demonstrated a wavelength-tunable metamaterial by tailoring magnetic resonance conditions with phase transition $VO_2$. The absorption peak shifts from 10.9 μm to 15.1 μm upon the $VO_2$ phase transition from metal to dielectric, resulting in a relative 38.5% shift in the peak wavelength. The underlying physical mechanisms lie in the plasmon-assisted magnetic resonance in metallic $VO_2$ and phonon-mediated counterpart in dielectric $VO_2$, which leads to different resonance wavelengths. A hybrid magnetic resonance mode due to phonon-plasmon coupling was also discussed when replacing the bottom $VO_2$ layer with a gold film. The insights and understanding gained in this work will facilitate the design of novel tunable metamaterials in active control of electronic, optical, and thermal devices.

This work was supported by the New Faculty Startup fund at Arizona State University.